\documentclass[letterpaper, 10 pt, conference]{ieeeconf}  

\IEEEoverridecommandlockouts                              
\usepackage{graphics} 
\usepackage{amsmath} 
\usepackage{amssymb}  
\usepackage{tikz}
\usepackage{pgfplots}
\usepackage{xcolor}
\usepackage{subcaption}
\usepackage{textcomp}
\usepackage{tabularx}

\usepackage[hidelinks]{hyperref}
\usepackage{url}

\usetikzlibrary{shapes,arrows}
\usetikzlibrary{arrows,shapes,positioning}
\usetikzlibrary{calc,decorations.markings}
\usetikzlibrary{spy}

\usetikzlibrary{shapes,arrows,calc,positioning,fit}

\usepackage{3dplot} 

\tikzset{surface/.style={draw=blue!70!black, fill=blue!40!white, fill
    opacity=.6}}

\definecolor{mycolor1}{rgb}{0.17860,0.52890,0.96820}%
\definecolor{mycolor2}{rgb}{0.85000,0.32500,0.09800}%

\newcommand{\frameI}{\mathrm{I}}
\newcommand{\frameB}{\mathrm{B}}

\newcommand{\pos}{\vec{p}}

\definecolor{gray10}{gray}{0.10}
\definecolor{gray20}{gray}{0.20}
\definecolor{gray30}{gray}{0.30}
\definecolor{gray40}{gray}{0.40}
\definecolor{gray50}{gray}{0.50}
\definecolor{gray60}{gray}{0.60}
\definecolor{gray70}{gray}{0.70}
\definecolor{gray80}{gray}{0.80}
\definecolor{gray90}{gray}{0.90}
\definecolorset{rgb}{}{}{myred,     0.7922,0.0000,0.1255}
\definecolorset{rgb}{}{}{myltred,   0.9569,0.6471,0.5098}

\definecolorset{rgb}{}{}{myltgreen, 0.6980,0.8745,0.5412}
\definecolorset{rgb}{}{}{mygreen,   0.2000,0.6275,0.1725}
	
\definecolorset{rgb}{}{}{myblue,    0.1216,0.4706,0.7059}
\definecolorset{rgb}{}{}{myltblue,  0.6510,0.8078,0.8902}

\definecolorset{rgb}{}{}{myorange,  0.9882,0.5529,0.3490}
	
\definecolorset{rgb}{}{}{myyellow,  0.9961,0.8784,0.5647}

\definecolorset{rgb}{}{}{mypurple,    0.4824,0.1961,0.5804}
\definecolorset{rgb}{}{}{myltpurple,  0.7608,0.6471,0.8118}

\definecolorset{rgb}{}{}{xaxiscolour,     0.7922,0.0000,0.1255}
\definecolorset{rgb}{}{}{yaxiscolour,   0.2000,0.6275,0.1725}
\definecolorset{rgb}{}{}{zaxiscolour,    0.1216,0.4706,0.7059}

\definecolorset{rgb}{}{}{forcecolour,     0.4824,0.1961,0.5804}
\definecolorset{rgb}{}{}{torquecolour,    0.4824,0.1961,0.5804}
\definecolorset{rgb}{}{}{gravitycolour,   0.4824,0.1961,0.5804}

\tikzset{ 
	table/.style={
		matrix of nodes,
		row sep=-\pgflinewidth,
		column sep=-\pgflinewidth,
		nodes={rectangle,text width=3em,align=center},
		text depth=1.25ex,
		text height=2.5ex,
		nodes in empty cells
	},
}

\pgfmathsetmacro{\yawintrinsic}{0}     
\pgfmathsetmacro{\pitchintrinsic}{0}
\pgfmathsetmacro{\rollintrinsic}{0}

\pgfmathsetmacro{\quaternioni}{0}
\pgfmathsetmacro{\quaternionj}{0}
\pgfmathsetmacro{\quaternionk}{0}
\pgfmathsetmacro{\quaternionr}{0}

\pgfmathsetmacro{\yawextrinsic}{0}     
\pgfmathsetmacro{\pitchextrinsic}{0}
\pgfmathsetmacro{\rollextrinsic}{0}

\pgfmathsetmacro{\rotmatIBoneone}{0}
\pgfmathsetmacro{\rotmatIBonetwo}{0}
\pgfmathsetmacro{\rotmatIBonethr}{0}
\pgfmathsetmacro{\rotmatIBtwoone}{0}
\pgfmathsetmacro{\rotmatIBtwotwo}{0}
\pgfmathsetmacro{\rotmatIBtwothr}{0}
\pgfmathsetmacro{\rotmatIBthrone}{0}
\pgfmathsetmacro{\rotmatIBthrtwo}{0}
\pgfmathsetmacro{\rotmatIBthrthr}{0}

\pgfmathsetmacro{\xIone}{0}
\pgfmathsetmacro{\yIone}{0}
\pgfmathsetmacro{\zIone}{0}

\pgfmathsetmacro{\xItwo}{0}
\pgfmathsetmacro{\yItwo}{0}
\pgfmathsetmacro{\zItwo}{0}

\pgfmathsetmacro{\xIthr}{0}
\pgfmathsetmacro{\yIthr}{0}
\pgfmathsetmacro{\zIthr}{0}

\pgfmathsetmacro{\xBone}{0}
\pgfmathsetmacro{\yBone}{0}
\pgfmathsetmacro{\zBone}{0}

\pgfmathsetmacro{\xBtwo}{0}
\pgfmathsetmacro{\yBtwo}{0}
\pgfmathsetmacro{\zBtwo}{0}

\pgfmathsetmacro{\xBthr}{0}
\pgfmathsetmacro{\yBthr}{0}
\pgfmathsetmacro{\zBthr}{0}

\newcommand{\computeRotationMatrixFromIntrinsic}[3]{
	
	\pgfmathsetmacro{\yawintrinsic}{#1}
	\pgfmathsetmacro{\pitchintrinsic}{#2}
	\pgfmathsetmacro{\rollintrinsic}{#3}
	
	\pgfmathsetmacro{\rotmatIBoneone}{
		+cos(\yawintrinsic)*cos(\pitchintrinsic)
	}
	\pgfmathsetmacro{\rotmatIBonetwo}{
		+cos(\yawintrinsic)*sin(\pitchintrinsic)*sin(\rollintrinsic)
		-sin(\yawintrinsic)*cos(\rollintrinsic)
	}
	\pgfmathsetmacro{\rotmatIBonethr}{
		+cos(\yawintrinsic)*sin(\pitchintrinsic)*cos(\rollintrinsic)
		+sin(\yawintrinsic)*sin(\rollintrinsic)
	}
	\pgfmathsetmacro{\rotmatIBtwoone}{
		+sin(\yawintrinsic)*cos(\pitchintrinsic)
	}
	\pgfmathsetmacro{\rotmatIBtwotwo}{
		+sin(\yawintrinsic)*sin(\pitchintrinsic)*sin(\rollintrinsic)
		+cos(\yawintrinsic)*cos(\rollintrinsic)
	}
	\pgfmathsetmacro{\rotmatIBtwothr}{
		+sin(\yawintrinsic)*sin(\pitchintrinsic)*cos(\rollintrinsic)
		-cos(\yawintrinsic)*sin(\rollintrinsic)
	}
	\pgfmathsetmacro{\rotmatIBthrone}{
		-sin(\pitchintrinsic)
	}
	\pgfmathsetmacro{\rotmatIBthrtwo}{
		cos(\pitchintrinsic)*sin(\rollintrinsic)
	}
	\pgfmathsetmacro{\rotmatIBthrthr}{
		cos(\pitchintrinsic)*cos(\rollintrinsic)
	}

}

\newcommand{\rotatepointoneIB}[3]{
	\pgfmathsetmacro{\xIone}{
		\rotmatIBoneone*#1 + \rotmatIBonetwo*#2 + \rotmatIBonethr*#3
	}
	\pgfmathsetmacro{\yIone}{
		\rotmatIBtwoone*#1 + \rotmatIBtwotwo*#2 + \rotmatIBtwothr*#3
	}
	\pgfmathsetmacro{\zIone}{
		\rotmatIBthrone*#1 + \rotmatIBthrtwo*#2 + \rotmatIBthrthr*#3
	}
}

\newcommand{\tikzgroundplane}[9]{
	
	\foreach \x in {0,...,#5}
		\draw[#9  , line width = #8 , - , #7 , tdplot_main_coords ] (#1+\x*#3,#2,0) -- (#1+\x*#3,#2+#4*#6,0);
		
	\foreach \y in {0,...,#6}
		\draw[#9  , line width = #8 , - , #7 , tdplot_main_coords ] (#1,#2+\y*#4,0) -- (#1+#3*#5,#2+\y*#4,0);
	
}

\DeclareMathOperator*{\argmin}{arg\,min} 

\newcommand{\lineWidth}{1.2}
\newcommand{\lineWidthB}{0.5}
\newcommand{\lineWidthC}{1.2pt}
\pgfplotsset{compat=1.11}

\newcommand*{\discountFactor}{\xi}

\title{\LARGE \bf
  Nonlinear Control of Quadcopters via Approximate Dynamic Programming
}

\author{
	Angel Romero, Paul N.\ Beuchat, Yvonne R.\ St\"urz, Roy S.\ Smith, and John Lygeros
	\thanks{All authors are with the Automatic Control Laboratory at the Swiss Federal Institute of Technology in Zurich (ETHZ)
		, Physikstrasse 3, 8092 Zurich, Swizerland. 
		Email addresses:
		\tt\small \{beuchatp, stuerzy, rsmith, jlygeros\}@ethz.ch}.
}

\begin{document}

\maketitle
\thispagestyle{empty}
\pagestyle{empty}

\begin{abstract}
While Approximate Dynamic Programming has successfully been used in many applications involving discrete states and inputs such as playing the games of Tetris or chess, it has not been used in many continuous state and input space applications. 
In this paper, we combine Approximate Dynamic Programming techniques and apply them to the continuous, non-linear and high dimensional dynamics of a quadcopter vehicle. 
We use a polynomial approximation of the dynamics and sum-of-squares programming techniques to compute a family of polynomial value function approximations for different tuning parameters. 
The resulting approximations to the optimal value function are combined in a point-wise maximum approach, which is used to compute the online policy. The success of the method is demonstrated in both simulations and experiments on a quadcopter. 
The control performance is compared to a linear time-varying Model Predictive Controller. 
The two methods are then combined to keep the computational benefits of a short horizon MPC and the long term performance benefits of the Approximate Dynamic Programming value function as the terminal cost. 
\end{abstract}


\section{INTRODUCTION}

%
Dynamic Programming (DP), \cite{Bellman1952}, is an important method for optimal decision making, relevant in a broad range of applications including finance, health sciences and engineering. While DP has been successfully applied to many specific problems \cite{DP_water}, it suffers from severe limitations when dealing with high dimensional systems. Approximate Dynamic Programming (ADP) aims to alleviate these limitations by finding tractable approximations to the optimal policies that result from DP. ADP methods have proven to work well for certain practical 
applications such as the game of Tetris \cite{Tetris}, or the game of chess \cite{Bellman_chess}. However, despite considerable progress in the theory and algorithms of ADP, \cite{DeFariasVanRoy,powell_2011_review,2004_adp_handbook}, there have not been many successful applications of ADP to systems with continuous state and input spaces.

%
In \cite{Wang}, the authors suggest the so-called \emph{iterated Bellman inequality} as a novel way of obtaining approximations to the value function by solving convex optimization problems. The approach was validated by applying it to the linear model of a simple mechanical system.
In \cite{Bartolomeo} the iterated Bellman inequality method is used to approximate the tail cost of a Model Predictive Control (MPC) problem of a continuous state space and finite input space power inverter. This effectively allows for a shorter time horizon, and thus a reduced computational burden.
In \cite{SOSummers}, the authors generalize the problem formulation of \cite{Wang} by considering polynomials as the chosen function space, and then use sum-of-squares (SOS) techniques to relax the problem to a semi-definite program (SDP). This method is implemented and applied to control a miniature-helicopter using a simplified linear model around hover.

%
The combination of ADP methods we apply to quadcopters is most closely connected with the method proposed in \cite{PaulPWMADP}. There the authors introduce a novel way of computing successively tighter under-estimators to the optimal value function in an iterative fashion.  This method has proven to be computationally tractable for the nonlinear quadcopter model considered in this paper, in contrast to \cite{Wang}.
In \cite{StateRelevanceWeighting}, a family of objective functions is chosen instead of only one, with a value function approximation computed for each and then put together in a point-wise maximum (PWM) approach. This reduces the dependency of the result on the choice of the individual objectives.
Parts of these techniques are used in \cite{Marc} and demonstrated in simulation on a nonlinear energy storage system with a low dimensional state-by-input space. In that application, the slow time scale allows for computationally demanding online policies.

%
Due to their inherent instability and ease of design, unmanned aerial vehicles (UAV) have widely been used as a test bench for control and robotics research, \cite{UAVUseCases}. In \cite{Minas}, nonlinear MPC is used to control the fast attitude dynamics of a hexacopter vehicle, and a Linear Quadratic Regulator (LQR) is used for the slower position dynamics. A nonlinear constraint on the tilt angle is considered. The maneuverability and tracking performance of the system are shown in simulation and experiment while tracking setpoints on a square. Similarly, in \cite{MPC_quadrotors}, a real time iteration linear MPC for the control of the position dynamics is run on-board a quadcopter.

%
In this paper, we combine multiple ADP methods to design a controller for the continuous, nonlinear, high dimensional model of a quadcopter and demonstrate the success of the methods in simulations and experiments. 
The key contributions of this work are: 
\begin{itemize}
	\item We combine the ADP techniques from \cite{SOSummers}, \cite{PaulPWMADP} and \cite{StateRelevanceWeighting}, leveraging the strengths of each.
	\item We develop the necessary approximation steps for synthesizing an ADP controller by this combined method for a high-dimensional quadcopter model. 
	\item We demonstrate the benefits of this method compared to other linear and nonlinear control techniques in simulations and experiments. 
\end{itemize}

The paper is organized as follows: In Section~\ref{sec:ADP_theory}, we present the deterministic control problem and introduce the ADP formulation. In Section~\ref{sec:quadcopter_control}, we derive the nonlinear model of the quadcopter, transform it to fit in our ADP formulation, and use it then to derive the approximation to the
optimal value function. Section~\ref{sec:numerical_results} presents the results obtained from simulations and experiments.


\section{Approximate Dynamic Programming Formulation}\label{sec:ADP_theory}

In this section, we first state the optimal control problem to be solved, and then we present the techniques used for computing approximate solutions, which will be applied to a quadcopter in the next section.

\subsection{Optimal Control Problem}

We consider the discrete-time dynamical system of a quadcopter with continuous state and input spaces and aim to minimize a discounted cost objective. The specific states, inputs and dynamics of the quadcopter will be given in Section~\ref{sec:quadcopter_control}. For now, let us denote the state of the system at time $k$ as $\smash{x_k \in \mathbb{R}^{n_x}}$, and the control input as $\smash{u_k \in \mathbb{R}^{n_u}}$. The evolution of the system is described by the dynamics function $\smash{f:\mathcal{X}\times\mathcal{U}\to\mathcal{X}}$ such that
\begin{equation}\label{eq:dynamics_markov}
	x_{k+1} = f(x_k,u_k).
\end{equation}
The stage cost function is the non-negative cost of taking decision $u_k$ when being in state
$x_k$, denoted as $\smash{l:\mathcal{X}\times\mathcal{U}\rightarrow\mathbb{R}_+}$. Given a \emph{discount factor} $\smash{\discountFactor \in [0,1)}$ and a stationary control policy of the form $\smash{\pi : \mathcal{X} \rightarrow \mathcal{U}}$, the discounted infinite horizon cost is thus,
\begin{equation}\label{eq:sum_stage_costs}
	J_\pi(x) = \sum_{k=0}^{\infty} \discountFactor^k \, l(x_k, \pi(x_k) ) ,\hspace{1.0cm} \text{with $x_0=x$.}
\end{equation}
The aim is to find the policy that minimizes \eqref{eq:sum_stage_costs} for all $x \in \mathcal{X}$.
The solution is characterized by the optimal value function $\smash{V^*:\mathcal{X}\to\mathbb{R}}$ that satisfies the Bellman equation,
\begin{equation}\label{eq:bellman_equation}
	V^*(x) \,=\, \underbrace{\,
			\min_{u \in \mathcal{U}}
			\hspace{0.1cm}
			l(x,u) + \discountFactor \, V^*(f(x,u))
		\,}_{\mathcal{T}V^*}\,,
		\hspace{0.5cm}\forall x \in \mathcal{X},
\end{equation}
where $\mathcal{T}$ is the Bellman operator. Given a solution of \eqref{eq:bellman_equation} the optimal policy is the so called \emph{greedy policy},
\begin{equation}\label{eq:greedy_policy_true}
	\pi^*(x) = \argmin_{u \in \mathcal{U}} \hspace{0.1cm} l(x,u) + \discountFactor \, V^*(f(x,u)).
\end{equation}
which minimizes the cost of taking the decision $u$ now, plus the cost-to-go from the next time period onwards.
We require the standard assumptions on the stage cost, dynamics, and existence of feasible policies to ensure that $V^*$ and $\pi^*$ exist, are time-invariant, and attain the minimum, for example \cite[Assumption 4.2.1, 4.2.2]{Lasserre} or \cite[\textsection 1.2]{Bertsekas2007}.
For a general problem instance, solving \eqref{eq:bellman_equation} and implementing \eqref{eq:greedy_policy_true} is intractable.

\subsection{Linear Programming Approach to ADP with Polynomials}\label{sec:LP_ADP}

The monotone and contractive properties of the Bellman operator mean that any function $\smash{\hat{V}:\mathcal{X}\to \mathbb{R}}$ satisfying the so called \emph{Bellman inequality} is a point-wise under-estimator of $V^*$, i.e.,
\begin{equation}\label{eq:bellman_inequality}
	\hat{V}(x) \leq \mathcal{T}\hat{V}(x),\, \forall x\!\in\!\mathcal{X}
		\hspace{0.1cm} \Rightarrow \hspace{0.1cm}
		\hat{V}(x) \leq V^*(x),\, \forall x\!\in\!\mathcal{X},
\end{equation}
where $\hat{V}$ is referred to as an \emph{approximate value function}.
This is the key property that motivates the various LP approaches to ADP proposed in \cite{DeFariasVanRoy,Wang,SOSummers,PaulPWMADP,Marc}. Letting $\smash{\mathcal{F(X)}}$ denote a space of functions mapping $\smash{\mathcal{X} \rightarrow \mathbb{R}}$, and using the Bellman inequality, a point-wise under-estimator is found by solving the following optimization problem,
\begin{subequations}\label{eq:opt_ADP}
	\begin{align} 
		\underset{\hat{V}(x)}{\text{maximize}} \hspace{0.2cm}
			& {\displaystyle\int  \hat{V}(x) \, c(x) \, dx}
			\label{eq:opt_ADP_1}
		\\
		\text{subject to} \hspace{0.2cm}
			& \;\;\hat{V} \in \mathcal{F(X)}
			\label{eq:opt_ADP_2}
		\\
		& \;\;\hat{V}(x) \leq \mathcal{T}\hat{V}(x), \hspace{0.1cm} \forall x \in
		\mathcal{X},
			\label{eq:opt_ADP_3}
	\end{align}
\end{subequations}
where the \emph{state relevance weighting} function, $c(x)$, is a finite measure on the state space and it allows the practitioner to select the region where a better approximation is desired. When $\mathcal{F(X)}$ is the space of real-valued measurable functions on $\mathcal{X}$ it is proven in \cite{Lasserre} that the solution to \eqref{eq:opt_ADP} solves the Bellman equation for $c$-almost all $\smash{x\in\mathcal{X}}$, however the infinite dimensional decision variable makes the problem intractable.

In this paper we use the space of polynomials for the function space $\mathcal{F(X)}$. This choice is motivated by \cite{SOSummers} where the authors show that problem \eqref{eq:opt_ADP} can be reformulated as a finite SDP when the dynamics, stage costs, and spaces are described by polynomials. Letting $\smash{\mathcal{P}_d\mathcal{(X)}}$ denote the space of polynomials up to degree $d$, we replace \eqref{eq:opt_ADP_2} by $\smash{\hat{V} \in \mathcal{P}_d\mathcal{(X)}}$ and hence the decision variables are the coefficients of the monomials up to degree $d$. Considering the minimization in the Bellman operator $\mathcal{T}$, the infinite constraints \eqref{eq:opt_ADP_3} can be equivalently written as,
\begin{equation}\label{eq:BI_relaxed}
	0 \leq \underbrace{ -\hat{V}(x) + l(x,u) + \discountFactor{\hat{V}(f(x,u))} \,}_{b(x,u)},
		\hspace{0.1cm} \forall \, x \!\in\!	\mathcal{X},\, u \!\in\! \mathcal{U},
\end{equation}
where $b(x,u)$ is a polynomial when $l$ and $f$ are polynomials.

We now use the SOS S-Procedure \cite{Parrilo} to reformulate \eqref{eq:BI_relaxed} as a single Linear Matrix Inequality (LMI) constraint on the decision variable $\smash{\hat{V} \in \mathcal{P}_d\mathcal{(X)}}$. Letting $SOS$ denote the set of polynomials that are representable as a sum of polynomials squared, then from \cite[Theorem 3.3]{Parrilo} we have the following equivalence for certifying that a polynomial $\smash{v \in \mathcal{P}_{2d}\mathcal{(X)}}$ is an SOS polynomial,
\begin{equation}
	v(x) \in SOS
		\hspace{0.15cm}\Leftrightarrow\hspace{0.15cm}
		\exists M \succeq 0,
		\hspace{0.1cm}\text{s.t.}\hspace{0.1cm}
		v(x) = z(x)^TMz(x),
\end{equation}
where $z(x)$ is the vector of monomials of $\smash{x\in\mathbb{R}^{n_x}}$ up to degree $d$, and $M$ is a square matrix of size ${n_x+d \choose d}$. Letting $\smash{g_i(x,u) \geq 0}$ denote the polynomials that describe the state and input spaces ($\mathcal{X}$ and $\mathcal{U}$), then the following set of equations,
\begin{subequations}\label{eq:SOS_procedure}
	\begin{align}
		b(x,u) - \lambda(x,u) \sum\nolimits_i g_i(x,u)
			&\,\in\, SOS,
			\label{eq:SOS_procedure_1}
		\\
		\lambda(x,u)
			&\,\in\, SOS,
			\label{eq:SOS_procedure_2}
	\end{align}
\end{subequations}
is a sufficient reformulation of \eqref{eq:BI_relaxed} in the sense that ${\text{\eqref{eq:SOS_procedure}} \Rightarrow \text{\eqref{eq:BI_relaxed}}}$.
Replacing \eqref{eq:opt_ADP_3} with \eqref{eq:SOS_procedure}, the multiplier $\smash{\lambda \in \mathcal{P}_d(\mathcal{X}\!\times\!\mathcal{U})}$ is an additional decision variable with the polynomial degree of the multiplier as a choice. It is reasonable to choose the largest multiplier degree for which the LMI constraints stemming from \eqref{eq:SOS_procedure} are computationally tractable.

Finally, the objective \eqref{eq:opt_ADP_1} is linear in the coefficients of $\smash{\hat{V} \in \mathcal{P}_d\mathcal{(X)}}$ and requires knowledge of the moments of the state relevance weighting function $c(x)$ up to order $d$. Thus, a point-wise under-estimator of $V^*$ can be found using a commercial solver for the SDP relaxation of \eqref{eq:opt_ADP}.

\subsection{Point-Wise Maximum Approach to ADP}\label{sec:PWM_ADP}

To improve the quality of the approximate value function, we use the approach proposed in \cite{PaulPWMADP} that solves a sequence of optimization problems, each with constraints of the same size as \eqref{eq:SOS_procedure}.

First, we solve the SDP relaxation of \eqref{eq:opt_ADP} for a particular choice of the state relevance weighting, and denote the solution as $\hat{V}_1^*$. Then we solve the SDP relaxation of the following optimization problem with $j=2$,
\begin{subequations}\label{eq:opt_LP_ADP_PWM}
	\begin{align}
	\underset{\hat{V}_j}{\text{maximize}} \hspace{0.2cm}
	& \displaystyle\int \hat{V}_j(x) \, c(x) \, dx
	\label{eq:opt_LP_ADP_PWM_1}
	\\
	\text{subject to} \hspace{0.2cm}
	& \hat{V}_j \in \mathcal{P}_d\mathcal{(X)},
	\label{eq:opt_LP_ADP_PWM_2}
	\\
	& \hat{V}_{j}(x) \leq \left( \mathcal{T}  \underset{k = 1, \dots, j-1}{\text{max}} \hat{V}^*_k \right) (x),
	\hspace{0.2cm}
	\forall x \!\in\! \mathcal{X}
	\label{eq:opt_LP_ADP_PWM_3}
	\end{align}
\end{subequations}
where $\hat{V}_1^*$ is fixed problem data, and we denote the solution as $\hat{V}_2^*$. We then solve the SDP relaxation of \eqref{eq:opt_LP_ADP_PWM} iteratively for ${j\geq 3}$, each time storing the solution $\hat{V}_j^*$ as fixed data to be used in the next iteration.
%
As the cost function \eqref{eq:opt_LP_ADP_PWM_1} is non-decreasing with the iterations of \eqref{eq:opt_LP_ADP_PWM}, we terminate when the improvement in the cost function becomes less than a pre-specified threshold. The approximate value function for a particular choice of $c(x)$ is set as the solution of the final iteration.

The steps given in \cite{PaulPWMADP} show how to reformulate \eqref{eq:opt_LP_ADP_PWM_3} as a polynomial inequality constraint similar to \eqref{eq:BI_relaxed}. The SOS S-Procedure is then applied and the resulting relaxation involves one LMI constraint with the same size as \eqref{eq:SOS_procedure_1}, and $\smash{j-1}$ LMI constraints identical to \eqref{eq:SOS_procedure_2}.

\subsection{Online Policy}\label{sec:tuning_sensitivity}

To construct an online policy, we first construct our best approximation to the value function, and we use this as a surrogate for $V^*$ in the greedy policy \eqref{eq:greedy_policy_true}. For different choices of the state relevance weighting, denoted $c_i(x)$ for $\smash{i=1,\dots,N_c}$, we perform the iteration described in Section \ref{sec:PWM_ADP} and denote $\hat{V}_{c_i}^*$ as the value function estimate returned by the final iteration. We take the PWM of these under-estimators as our best estimate for $V^*$, i.e., 
\begin{equation} \label{eq:Vbest}
		\max_{i = 1,\dots, N_c} \hspace{0.1cm} \hat{V}_{c_i}^*(x)
		\hspace{0.2cm}\leq\hspace{0.2cm}
		V^*(x),
		\hspace{0.2cm} \forall x \in \mathcal{X}.
\end{equation}
The online policy, also called the \emph{approximate greedy policy}, is thus,
\begin{equation}\label{eq:greedy_policy_ADP_PWM}
	\begin{aligned}
		\hat{\pi}(x)
		&=\argmin_{u \in \mathcal{U}} \hspace{0.1cm} l(x,u) + \discountFactor\, \max_{i = 1,\dots, N_c} \hspace{0.1cm} \hat{V}_{c_i}^*(f(x,u)),
	\end{aligned}
\end{equation}
where the maximum over the $\hat{V}_{c_i}^*$ is readily reformulated using an epigraph variable.

To construct the surrogate for $V^*$, it remains to choose the $c_i$ weightings and in Section \ref{sec:numerical_results} we provide the weightings used for nonlinear control of a quadcopter.
The surrogate for $V^*$ can be considered as a sufficiently accurate estimate if the online performance of \eqref{eq:greedy_policy_ADP_PWM} achieves a low cost.


\section{Control of a Quadcopter}\label{sec:quadcopter_control}

In the following, the theory from the previous section will be used for the control of a quadcopter vehicle. The dynamics are brought into a suitable form by applying a reduction in the state space and a polynomial approximation.

\subsection{Dynamical Model of the Quadcopter}
To describe the dynamics of the quadcopter, two frames of reference are used, the \emph{inertial (world) frame} and the \emph{body frame} (attached to the quadcopter). The location of the body frame with respect to the inertial frame is denoted as ${\vec{p} = [p_x, p_y, p_z]^\top}$ and shown in Figure~\ref{fig:quadcopter}. 
The angular rates about the body frame axes $x^{(B)}$, $y^{(B)}$ and $z^{(B)}$ are denoted ${\vec{\omega} = [w_x, w_y, w_z]^\top}$. The orientation of the body frame relative to the inertial frame is given by ${\vec{\psi} = [\gamma,\beta,\alpha]^\top}$, the roll, pitch, and yaw intrinsic Euler angles, respectively. Using the ZYX convention for the Euler angles, and with ${s_{(\cdot)}:=\sin(\cdot)}$, ${c_{(\cdot)}:=\cos(\cdot)}$, the equations of motion for a quadcopter of mass $m$ and with moment of inertia $J$, are readily derived as,
\begin{subequations}\label{eq:newton_dynamics}
	\begin{align}
		\ddot{\vec{p}}^{(I)} &= \frac{1}{m}\sum\limits_{i = 1}^{4}
		f_i
		\begin{bmatrix}
		c_\alpha s_\beta s_\gamma + s_\alpha c_\gamma\\
		s_\alpha s_\beta c_\gamma - c_\alpha s_\gamma\\
		c_\beta c_\gamma
		\end{bmatrix} -
		\begin{bmatrix}
		0\\
		0\\
		g
		\end{bmatrix}\label{eq:newton_dynamics_1},
		\\
		\begin{bmatrix}
		\dot{\omega_x}\\
		\dot{\omega_y}\\
		\dot{\omega_z}
		\end{bmatrix}
		&= J^{-1} \left(
		\begin{bmatrix}
		\sum_{i = 1}^{4} f_i d_{y_i}\\
		\sum_{i = 1}^{4} -f_i d_{x_i}\\
		\sum_{i = 1}^{4} f_i d_{\tau_i}
		\end{bmatrix}
		- \vec{\omega} \times J \vec{\omega} \right)\label{eq:newton_dynamics_2},
	\end{align}
\end{subequations}
where $f_i$ is the thrust produced by propeller $i$, $d_{y_i}$ and $d_{x_i}$ are the distances from the axis of propeller $i$ to the center of gravity of the quadcopter, $d_{\tau_i}$ is a constant of proportionality for how much torque is produced by thrust $f_i$, and $g$ is the acceleration due to gravity.
For more details on the derivation of the dynamics refer to \cite{Mahony}.
The system thus has 12 states, given by $[\vec{p}^{\,\top}, \dot{\vec{p}}^{\,\top}, \vec{\psi}^{\,\top}, \vec{\omega}^{\,\top}]^\top$, and four inputs, given by the motor thrusts $f_i$, $i=1,...,4$. %

\input{img_for_arXiv_v1/quadrotor_tikz.tex}

\subsection{System Model Approximation for ADP}

%
%

\begin{figure}
	\centering

\tikzstyle{sum} = [draw, fill=blue!20, circle, node distance=1cm]
\tikzstyle{input} = [coordinate]
\tikzstyle{output} = [coordinate]
\tikzstyle{pinstyle} = [pin edge={to-,thin,black}]

\begin{tikzpicture}[auto, node distance=2cm,>=latex', scale=0.6]
	\node [input, name=ref] {};
	\node [input, above = 5em of ref](f_T) {};
	\node [sum, right = 5em of ref](sum1) {};
	
	\tikzstyle{block} = [draw, fill=red!20, rectangle, 
	minimum height=3em, minimum width=2em]
	
	\node [block, right = 1.0em of sum1 , align=center]
	(outer_controller)
	{ \scriptsize{Outer}\\ \scriptsize{Controller} };

	\node [sum,right = 3.5em of outer_controller]
	(sum2){};

	\tikzstyle{block} = [draw, fill=violet!20, rectangle, 
	minimum height=3em, minimum width=3em]
	
	\node [block, right = 1.0em of sum2 , align=center]
	(inner_controller)
	{ \scriptsize{Inner}\\ \scriptsize{Controller} };

	\tikzstyle{block} = [draw, fill=blue!10, rectangle, minimum height=3em, minimum width=3em]
	
	\node [block, right = 3.2em of inner_controller, align=center]
	(dynamics)
	{\scriptsize{Plant}\\ \scriptsize{as per \eqref{eq:newton_dynamics}}};
	
	\node [output, right=1.5em of dynamics] (state_output) {};

	\draw [draw,->] (sum1) -- node [pos=0.3] {} (outer_controller);

	\draw [draw,->] (outer_controller) -- ++ (0,+1.7) -|
	node [above , pos=0.18] {\scriptsize{$f_{T,\text{ref}}$}}
	node [pos = 0.5,name=ft]{}
	(inner_controller);

	\draw [draw,->] (outer_controller) --
	node {\scriptsize{$\vec{\psi}_{\text{ref}}$}}
	(sum2);
	\draw [draw,->] (sum2) -- node {} (inner_controller);

	\draw [draw,->] (inner_controller) --
	node [above] {\scriptsize{$f_1,f_2$}}
	node[below]{\scriptsize{$f_3,f_4$}}
	(dynamics);
	
	\draw [draw,-] (dynamics) --
	(state_output);

	\draw [draw,->] (state_output) -- ++ (0,-1.7) -|
	node [pos=0.99] {$$}
	node [ above, pos=0.25]
	{\scriptsize{$\vec{\omega}$}}
	(inner_controller);
	\draw [draw,->] (state_output) -- ++ (0,-2.8) -|
	node [pos=0.99] {$-$}
	node [ above, pos=0.25, name=w]
	{\scriptsize{$\vec{\psi}$}}
	(sum2);

	\draw [draw,->] (state_output) -- ++ (0,-3.9) -|
	node [pos=0.99] {$-$}
	node [ above, pos=0.25,name=psi]
	{\scriptsize{$(\vec{p},\dot{\vec{p}})$}}
	(sum1);
	
\end{tikzpicture}
	\caption{
		Cascaded control structure. The references $f_{T,ref}$ and $\vec{\psi}_{ref}$ are the inputs $u$ used in the model for synthesizing the outer controller. The inner loop consists of PID controllers that track these attitude and thrust references.
      	\label{fig:control_arch}
	}
\end{figure}
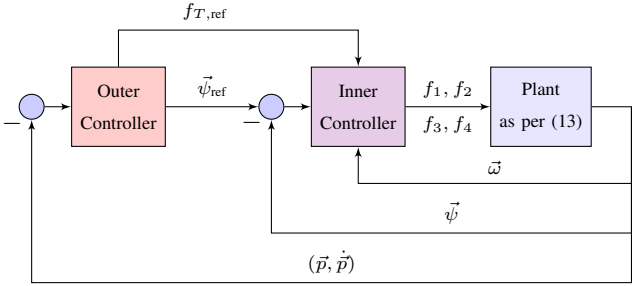

In order to apply the ADP method from Section~\ref{sec:ADP_theory}, we approximate the full dynamics of \eqref{eq:newton_dynamics} in three steps.
%
First, we exploit time scale separation to reduce the state dimension of the dynamics by using the cascaded control structure shown in Figure~\ref{fig:control_arch}.
%
The inner controller considers the states $[\vec{\psi\,}^\top,\vec{\omega}^\top]^\top$ and uses the thrust inputs $f_1,\dots,f_4$ to track a reference attitude $\vec{\psi}_{\mathrm{ref}}$ and total thrust $f_{T,\mathrm{ref}}$.
%
The outer controller considers the states ${x=[\,\vec{p}^\top,\dot{\vec{p}}^\top\,]^\top}$ and uses the inputs ${u=[\gamma,\beta,\alpha,f_T]^\top}$ to stabilize the quadcopter to a specified position and yaw setpoint. These inputs are given as references to the inner controller.
%
This cascaded structure is common practice for quadcopter control \cite{Minas}, and the approximation step assumes a sufficient time-scale separation so that the transients of the inner loop controller can be neglected when designing the outer loop controller.
%
The goal is to design the outer controller using the combination of ADP methods described in Section~\ref{sec:ADP_theory}.

In the second approximation step, we use a Taylor expansion of \eqref{eq:newton_dynamics_1} to replace the trigonometric functions with a polynomial approximation that fits the SOS framework described in Section~\ref{sec:LP_ADP}. We use a third order Taylor expansion as a compromise between the accuracy of the approximated nonlinear dynamics and the size of the optimization problem~\eqref{eq:opt_LP_ADP_PWM} for fitting approximate value functions. The translational dynamics~\eqref{eq:newton_dynamics_1} are thus approximated as

\vspace*{-0.3cm}
     
\begin{footnotesize}
	\begin{equation}\label{eq:dynamics_taylor_3_continous}
		\begin{bmatrix}
			\dot{\vec{p}} \\
			\frac{1}{m}(\beta f_{T}  -
			\frac{1}{2}mg\alpha^2\beta + f_{T}\alpha \gamma - \frac{1}{6}mg
			\beta^3 - \frac{1}{2}mg\beta \gamma^2 )\\
			\frac{1}{m}(-\gamma f_{T} + f_{T}\alpha\beta
			+\frac{1}{2}mg\alpha^2\gamma +\frac{1}{6}mg\gamma^3)\\
			\frac{1}{m}((f_{T} - mg) - \frac{1}{2}f_{T}\beta^2 -
			\frac{1}{2}f_{T}\gamma^2)\\
		\end{bmatrix},
	\end{equation}
\end{footnotesize}
where we have introduced ${f_T=\sum_i f_i}$ to represent the total thrust.

Finally, to bring the dynamics of \eqref{eq:dynamics_taylor_3_continous} to discrete time, we use a forward Euler scheme. The discrete-time dynamics are required for applying the ADP methods to the design of the outer-loop controller. We choose a forward Euler approximation in order not to increase the polynomial order of the dynamics, and assume that the sampling time is fast enough to maintain stability properties of the plant. 
For the stage cost we chose the quadratic form, ${l(x,u) = x^\top Qx + u^\top Ru}$.

\subsection{Input Constraints}


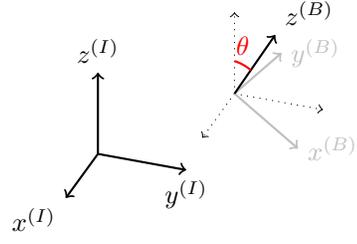
\begin{figure}
	\centering
	
	\tdplotsetmaincoords{60}{110}
	
	\pgfmathsetmacro{\rvec}{1}
	\pgfmathsetmacro{\thetavec}{30}
	\pgfmathsetmacro{\phivec}{80}
	
	\begin{tikzpicture}[scale=2.5,tdplot_main_coords]
		
		\coordinate (O) at (0,0,0);

		\coordinate (P) at (-0.2,0.7,0.4);
			
		
		\draw[thick,->] (0,0,0) -- (0.5,0,0) node[anchor=north east]{$x^{(I)}$};
		\draw[thick,->] (0,0,0) -- (0,0.5,0) node[anchor=north]{$y^{(I)}$};
		\draw[thick,->] (0,0,0) -- (0,0,0.5) node[anchor=south]{$z^{(I)}$};

		\tdplotsetrotatedcoords{0}{0}{0}
		
		\tdplotsetrotatedcoordsorigin{(P)}
		
		\draw[dotted,tdplot_rotated_coords,->] (0,0,0) -- (0.5,0,0);
		\draw[dotted,tdplot_rotated_coords,->] (0,0,0) -- (0,0.5,0);
		\draw[dotted,tdplot_rotated_coords,->] (0,0,0) -- (0,0,0.5);

		\tdplotsetrotatedcoords{\phivec}{\thetavec}{0}
		
		\draw[thick, color=black!25, tdplot_rotated_coords,->] (0,0,0) -- (.45,0,0) node[anchor=west]{$x^{(B)}$};
		\draw[thick, color=black!25, tdplot_rotated_coords,->] (0,0,0) -- (0,.5,0) node[anchor=west]{$y^{(B)}$};
		\draw[thick,tdplot_rotated_coords,->] (0,0,0) -- (0,0,.5)
		node[anchor=south west]{$z^{(B)}$};
		
		\tdplotsetthetaplanecoords{\phivec}
		
		\tdplotdrawarc[thick, tdplot_rotated_coords,color=red]{(P)}{0.2}{0}{\thetavec}{anchor=south,
		  color=red}{$\theta$}
		
	\end{tikzpicture}
	\caption{Visualization of the constraint on the tilt angle $\theta$. \label{fig:angle_constraint}}
\end{figure}

We chose to constrain the tilt angle that the $z^{(I)}$ axis forms with the $z^{(B)}$ axis, denoted by $\theta$ and shown in Figure~\ref{fig:angle_constraint}. The motivation behind this choice is linked to the nonlinearities in the dynamics. We choose $\theta$ large enough so that the nonlinearities are relevant for the solution of the optimal control problem, but small enough so that the third order Taylor expansion \eqref{eq:dynamics_taylor_3_continous} is an adequate approximation of the true dynamics \eqref{eq:newton_dynamics_1}. The angle constraint is given by 
\begin{equation}\label{eq:constraints_angle}
	\cos(\theta) \leq \cos(\beta)\cos(\gamma),
\end{equation}
which is nonlinear in the inputs $\gamma$ and $\beta$, similar to the constraint introduced in \cite{Minas}.

To represent the physical actuator limits of the propellers, we impose a lower and upper bound on the total thrust, $f_T$. Denoting the lower and upper bound as $f_{T,lb}$ and $f_{T,ub}$, respectively, the constraint is thus,
\begin{equation}\label{eq:constraints_f_T}
	f_{T,lb} \leq f_T \leq f_{T,ub}.
\end{equation}

In addition to the dynamics, the constraints also need to be in polynomial form in order to apply the SOS techniques from Section~\ref{sec:LP_ADP}. We chose quadratic forms for the constraints in \eqref{eq:constraints_angle} and \eqref{eq:constraints_f_T}.
The constraint on the angle in \eqref{eq:constraints_angle} is approximated via least-squares to the closest ellipse with radii $a_1$ and $a_2$, and the constraint on the total thrust in \eqref{eq:constraints_f_T} can be transformed into a single second order polynomial to reduce the number of SOS multipliers.
The constraint functions are thus given by 
%
\begin{subequations}\label{eq:constraints_quad}
	\begin{align} 
		1 - \frac{\gamma^2}{{a_1}^2} - \frac{\beta^2}{{a_2}^2} =: g_\theta(u)
			&\geq 0,\label{eq:constraints_angle_quad}
		\\
		f_T\left(f_{T,lb} + f_{T,ub} \right) - f_{T,lb}f_{T,ub} - f_T^2 =: g_f(u)
			&\geq 0.\label{eq:constraints_f_T_quad}
	\end{align}
\end{subequations}
%
Note that $g_f(u)$ and $g_\theta(u)$ correspond to $g_i(x,u)$ in \eqref{eq:SOS_procedure}.

\subsection{Value Function Fitting}\label{sec:Vfunc_fitting}

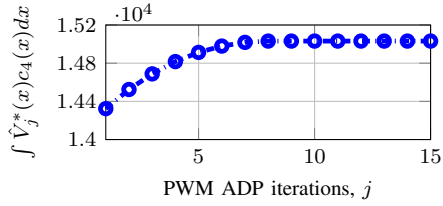
\begin{figure}
	\centering
%
%
%
\begin{tikzpicture}

\begin{axis}[%
width=1.7in,
height=0.6in,
at={(0.26in,0.1887in)},
scale only axis,
xmin=1,
xmax=15,
xlabel style={font=\color{white!15!black}, font=\footnotesize},
xlabel={{PWM ADP iterations, $j$}},
font=\footnotesize, ytick={14000, 14400, 14800,15200},
ymin=14000,
ymax=15200,
ylabel style={font=\color{white!15!black}, font=\footnotesize},
ylabel={$\int \hat{V}^*_j(x) c_4(x) dx$},
axis background/.style={fill=white},
tick label style={/pgf/number format/fixed, font=\footnotesize},
xmajorgrids,
ymajorgrids,
legend style={legend cell align=left, align=left, draw=white!15!black}
]
\addplot [color=blue, dashdotted, line width=1.8, mark=o, mark options={solid, blue}]
  table[row sep=crcr]{%
1	14323.5889157498\\
2	14524.3599895415\\
3	14687.9298721557\\
4	14816.6314518187\\
5	14912.8882651123\\
6	14979.1676980477\\
7	15017.9224101483\\
8	15031.5521366575\\
9	15031.5591179874\\
10	15031.5576383743\\
11	15031.5551923825\\
12	15031.5568230948\\
13	15031.5548437242\\
14	15031.5522758841\\
15	15031.5579090874\\
};

\end{axis}
\end{tikzpicture}%
	\caption{Convergence of the objective function value of \eqref{eq:opt_LP_ADP_PWM} over iterations $j$\label{fig:objective_evolution}, for state relevance weighting $c_4$ in  \eqref{eq:different_cs}.}
\end{figure}

\begin{figure}
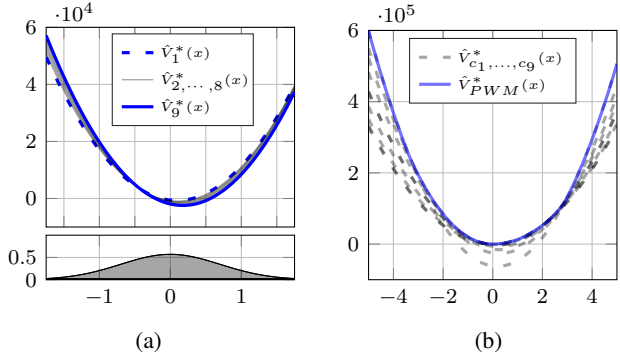

	\begin{subfigure}[t]{0.46\columnwidth}
	\input{img_for_arXiv_v1/tikz_z_dot}
	\caption{}
	\end{subfigure}
	\hspace*{0.3cm}
	\begin{subfigure}[t]{0.46\columnwidth}
	\input{img_for_arXiv_v1/tikz_z_dot_PWM}
	\caption{}
	\end{subfigure}
	\caption{
		Cuts around $\dot{p}_z$ \textbf{(a)}: of the approximate value functions $\hat{V}_j^*(x)$ for different iterations $j=1,...,9$ of \eqref{eq:opt_LP_ADP_PWM} for $c_4(x)$ as shown in the lower plot, \label{fig:iterations_z_dot} and \textbf{(b)}: of the value function approximations $\hat{V}_{c_i}^*(x)$ for the $c_i(x)$ in \eqref{eq:different_cs} and their PWM $\hat{V}_{PWM}^*(x)$.\label{fig:PWM_y_dot}
	}
\end{figure}

Heuristics have been developed for how to choose the state relevance weighting, $c(x)$, which is the main tuning parameter. 
As presented in \cite{StateRelevanceWeighting}, to alleviate the sensitivity of this choice, we select a family of $\smash{c_i(x) \sim \mathcal{N}(0,\Sigma_i)}$ as Gaussians with zero mean and different covariance matrices.
Let $\mathrm{diag}(v)$ define a diagonal matrix with the entries of vector $v$ on its diagonal. Then, with ${\Sigma_0 = \mathrm{diag}([0.1,0.1,0.1,1,1,1])}$, the set of covariance matrices $\Sigma_i$ used is obtained as,
\begin{equation}\label{eq:different_cs}
	\begin{aligned}
		&\Sigma_i = K_i \, \Sigma_0, ~i=\{1,...,9\}, ~ \text{with}\\
		&K_i \in \{0.01,\, 0.1, \,0.3, \,0.5, \,0.7, \,1, \,1.3, \,1.5, \,2\}.
	\end{aligned}
\end{equation}
For each different $c_i(x)$ in \eqref{eq:different_cs}, we perform the iterative procedure described in Section~\ref{sec:PWM_ADP}. 
As shown in Figure~\ref{fig:objective_evolution}, the relative cost improvement converges to under a threshold of $10^{-5}$ within 9 iterations.
In Figure~\ref{fig:iterations_z_dot}\,(a), a cut of the value function approximation for different iterations for the same $c_i(x)$ is shown. The  cut is obtained by setting all states to zero except $\dot{p}_z$.
Solving this iterative problem for the different $c_i(x)$ results in a family of value function approximations, $\hat{V}_{c_i}^*(x)$, $i=1,...,N_c$. As presented in Section~\ref{sec:tuning_sensitivity}, the point-wise maximum (PWM) of this family, ${\hat{V}^*_{PWM}(x) := \underset{i = 1, \dots, N_c}{\text{max}} \hat{V}^*_{c_i}(x)}$, is the value function approximation that is used for the online greedy policy \eqref{eq:greedy_policy_ADP_PWM}. 
Figure~\ref{fig:PWM_y_dot}\,(b) shows a cut in the $\dot{p}_z$ state of the value function approximations $\hat{V}_{c_i}^*(x)$ and of $\hat{V}^*_{PWM}(x)$.


\section{Numerical Results}\label{sec:numerical_results}
In this section, in order to evaluate the control performance of the presented ADP approach, we compare different control schemes, which will be explained in detail in the following, with an overview given in Table~\ref{tab:policynames}. The names are comprised of three parts, the first part encodes the approximation of the system dynamics used in the online policy, the middle part refers to the type of policy, and the last part indicates the terminal cost or value function used. 
The control task is a setpoint change of $2$ meters in $p_x$-direction, which shows the performance benefits of the control methods using the ADP value function compared to the other control schemes. 
First, results from simulations are given and then, experimental results on the quadcopter are presented.

\renewcommand{\arraystretch}{1.3}
\begin{table}
	\begin{center}
		\caption{Control policies compared in Section~\ref{sec:numerical_results}.}
		\label{tab:policynames}
		\begin{tabular}{p{19mm} p{20mm} p{13mm} p{11mm}}
			\hline
			\textbf{Name} &  \textbf{dynamics approximation} & \textbf{policy} & \textbf{terminal cost}
			\\
			[0.03cm]\hline
			NL-GP-ADP
			& 3rd order Taylor as in \eqref{eq:dynamics_taylor_3_continous}
			& Greedy policy \eqref{eq:greedy_policy_ADP_PWM}
			& $\hat{V}^\ast_{PWM}$,  \textsection\ref{sec:Vfunc_fitting}
			\\[0.03cm]
			\hline
			LTV-MPC-LQR & LTV as in \eqref{eq:MPC_problem} & MPC & $P$ from Riccati
			\\[0.03cm]
			\hline
			LTI-MPC-LQR & linearized around hover & MPC   & $P$ from Riccati
			\\[0.03cm]
			\hline
			LTV-MPC-ADP
			& LTV as in \eqref{eq:optimization_problem_general_ADP}
			& MPC
			& $\hat{V}^\ast_{PWM}$,  \textsection\ref{sec:Vfunc_fitting}
			\\[0.03cm]
			\hline 
		\end{tabular}
	\end{center}
\end{table}
\renewcommand{\arraystretch}{1.0}

\subsection{Simulation Results}
\label{sec:numerical_results_sim}

We consider the model of a small-sized quadcopter, the \emph{Crazyflie 2.0}, \cite{crazy}, with a mass of 27 $g$. Details about the model parameters such as its inertia and dimensions can be found in \cite{Angel}, \cite{dfallgit}. 
For the simulation the full state dynamics of the quadcopter as in \eqref{eq:newton_dynamics} are used, together with the control structure shown in Figure~\ref{fig:control_arch}.
%
The inner loop tracks the commanded reference Euler angles $\psi_{ref}$ with a PID controller running at 500\,Hz, with the controller parameters and implementation details found in \cite{dfallgit}.  The outer control loop, and hence all the controllers listed in Table \ref{tab:policynames}, runs at 50\,Hz.

In the ADP approach, the matrices for the stage cost, ${l(x,u) = x^\top Qx + u^\top Ru}$, have been chosen as,
\begin{equation}\label{eq:Q_R}
	\begin{aligned}
		Q &= \mathrm{diag}(\begin{bmatrix}50&50&50&5&5&5\end{bmatrix}),
		\\
		R &= \mathrm{diag}(\begin{bmatrix}20&20&100&1500\end{bmatrix}),
	\end{aligned}
\end{equation}
and the discount factor is set to $\discountFactor = 0.98$.
The total thrust is constrained to ${f_T\!\in\![0N,\,0.56N]}$.
For the angle constraint in \eqref{eq:constraints_angle} we have chosen $\theta = 45^{\circ}$, which is large enough to demonstrate that the controller takes into account the nonlinearities of the system, but small enough for the constraint to become active.

At each time step, the online policy in \eqref{eq:greedy_policy_ADP_PWM} is solved and the optimal input is applied to the system. As the term $\hat{V}_{c_i}^*(f(x,u))$ is a composition of a quadratic and a third order polynomial, the resulting  optimization problem is not convex. We refer to this as NL-GP-ADP, see Table~\ref{tab:policynames}, and we solve it in simulation with an interior point method.

In the following, we compare NL-GP-ADP to a linear time-varying (LTV) MPC, \cite{RTI}, which also accounts for the nonlinear dynamics in its predictions. 
At each time step, the dynamics \eqref{eq:newton_dynamics_1} are linearized around the predicted trajectory, $x_{\mathrm{traj},k}$, $u_{\mathrm{traj},k}$, and denoted by $A_k, B_k, g_k$. Then, the following problem is solved, 
\vspace*{-0.5cm}

\begin{small}
	\begin{equation}\label{eq:MPC_problem}
	\begin{aligned}
	& \underset{\{u_k\}_{k=0}^{N-1}}{\text{minimize}} & & \sum_{k=0}^{N-1} \lVert
	x_k - x_{\mathrm{traj}, k} \rVert_{Q}^{2} + \lVert u_k - u_{\mathrm{traj}, k}
	\rVert_{R}^{2}\\[-0.3cm]
	&&& \hspace{2.75cm} + \lVert x_N - x_{\mathrm{traj}, N} \rVert_{P}^{2}\\
	& \text{subject to } & &   x_{k+1} = A_kx_k + B_ku_k + g_k,\\
	&  & &   u_{k} \in \mathcal{U} ,\\
	\end{aligned}
	\end{equation}
\end{small}
where the set $\mathcal{U}$ is defined as in \eqref{eq:constraints_quad}, making this problem a convex Quadratically Constrained Quadratic Program (QCQP). Note that the $x_k$ and $u_k$ variables are deviations from the predicted trajectory. The cost matrices $Q$ and $R$ are the same as in \eqref{eq:Q_R}, and $P$ comes from the solution to the Riccati equation using the linearization of dynamics \eqref{eq:newton_dynamics_1} around the hover state ${[\,\vec{p},\dot{\vec{p}}\,]^\top}\!=\!0$. The first input, $u_0^*$, is applied, and the system evolves by one time step. The predicted trajectory for the next iteration, i.e., $x_{\mathrm{traj},k}$, $u_{\mathrm{traj},k}$, is based on the state measurement and the inputs $u_1^*,...,u_{N-1}^*$. We refer to this as LTV-MPC-LQR of horizon length $N$, see Table~\ref{tab:policynames}.

%
Furthermore, a comparison of the NL-GP-ADP policy to a controller that does not account for the nonlinear dynamics is given. For this we use a linear time-invariant (LTI) MPC, i.e., for the dynamics we use the linearized model around hover and the same $Q$, $R$, and $P$ cost and Riccati solution matrices as in \eqref{eq:MPC_problem}.
%
Consistent with the naming convention, this policy is referred to as LTI-MPC-LQR of horizon $N=1$, and the policy is also a convex QCQP.
%
This policy is similar to the ADP approach demonstrated on miniature helicopters in \cite{SOSummers}, and for the control task we consider the performance was almost identical. Thus, for the sake of clarity, we do not include a policy based on \cite{SOSummers} in the comparison here.

\begin{figure}
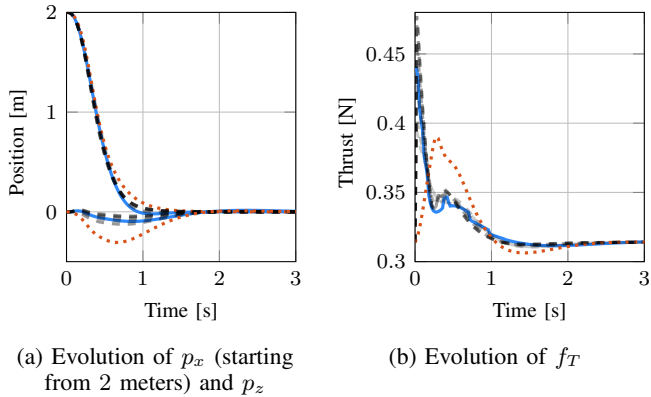

	\begin{subfigure}[t]{0.48\columnwidth}
		\centering \input{img_for_arXiv_v1/x_z_MPC_ADP}
		\caption{\centering Evolution of $p_x$ (starting from $2$ meters) and $p_z$}
	\end{subfigure}
	\hspace{0.0cm}
	\begin{subfigure}[t]{0.48\columnwidth}
		\centering \input{img_for_arXiv_v1/thrust_MPC_ADP}
		\caption{\centering Evolution of $f_T$}
	\end{subfigure}
	\caption{
		Comparison of the behavior of the system in simulation for: \ref{NADP} NL-GP-ADP, \ref{LADP} LTI-MPC-LQR ${N\!=\!1}$, and \ref{MPC} LTV-MPC-LQR of different horizon lengths (from lighter to darker, $\smash{N=1, 2, 5, 10, 15, 20}$).
		\label{fig:comparison_simulation}
	}
\end{figure}

A comparison between the three policies is shown in Figure~\ref{fig:comparison_simulation}.
The main difference between the approaches is the drop in $p_z$ due to the fact that NL-GP-ADP and LTV-MPC-LQR have more information about the nonlinearities of the system than LTI-MPC-LQR which uses the linearized dynamics around hover. The nonlinearity of interest for this test is the change in vertical thrust when the quadcopter tilts, and as seen in Figure~\ref{fig:comparison_simulation}\,(b), LTI-MPC-LQR corrects for the drop in $p_z$ after it happens. However,  NL-GP-ADP and LTV-MPC-LQR pre-compensate for this drop by increasing the thrust in advance.

\subsection{Experimental Results}

For the experiments, the Crazyflie quadcopter 
is used together with a set of infrared 
cameras that provide the system with position and attitude measurements, \cite{vicon}. Body frame angular rate measurements come from the on-board IMU. The control loops are the same as described in Section~\ref{sec:numerical_results_sim} for the simulations. 

Implementation of the NL-GP-ADP in real-time (i.e., at 50\,Hz) was computationally intractable. To overcome this difficulty, the ADP approach is combined with the linear time-varying MPC scheme by replacing the terminal cost in \eqref{eq:MPC_problem} with the ADP value function approximation, leading to 
\vspace{-0.4cm}

\begin{small}
	\begin{equation}
	\begin{aligned}\label{eq:optimization_problem_general_ADP}
	&\underset{\{u_k\}_{k=0}^{N-1}}{\text{minimize }}
	\hspace{0.1cm}
	&&\sum_{k=0}^{N-1} \lVert
	x_k - x_{\mathrm{traj}, k} \rVert_{Q}^{2} + \lVert u_k - u_{\mathrm{traj}, k}
	\rVert_{R}^{2}
	\\[-0.3cm]
	&&& \hspace{2.75cm} +
	\hat{V}^*_{PWM}(x_N - x_{\mathrm{traj},N})
	\\
	& \text{subject to } & &   x_{k+1} = A_kx_k + B_ku_k + g_k,\\
		&  & &   u_{k} \in \mathcal{U}. \\[-0.3cm]
	\end{aligned}
	\end{equation}
\end{small}
\vspace{-0.2cm}

\noindent We call this the LTV-MPC-ADP policy of horizon $N$. 
If we introduce the epigraph of all quadratics $\hat{V}^*_{c_i}(x)$ in $\hat{V}^*_{PWM}$ as a new optimization variable, as in \cite{StateRelevanceWeighting}, the resulting optimization problem is a QCQP, making problem \eqref{eq:optimization_problem_general_ADP} implementable in real-time.
Particularly, if in \eqref{eq:optimization_problem_general_ADP} we choose the horizon length to be 1, we have the \emph{linearized greedy policy}, which differs from problem~\eqref{eq:greedy_policy_ADP_PWM} in that the prediction of the next state is done using the linearized dynamics around the current state, instead of the nonlinear dynamics.

\begin{figure*}
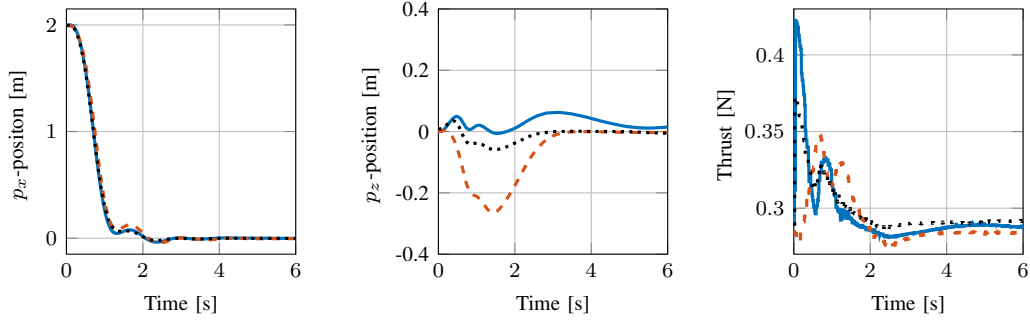

	\hspace{2.0cm}
	\begin{subfigure}[t]{0.25\textwidth}
		\input{img_for_arXiv_v1/x_impl_MPCADP_MPC}
	\end{subfigure}~
	%
	%
	\begin{subfigure}[t]{0.25\textwidth}
		\input{img_for_arXiv_v1/z_impl_MPCADP_MPC}
	\end{subfigure}~
	\begin{subfigure}[t]{0.25\textwidth}
		\centering
		\input{img_for_arXiv_v1/thrust_impl_MPCADP_MPC}
	\end{subfigure}
	\caption{ 
		Experimental comparison of different policies: \ref{MPCADP1} LTV-MPC-ADP, \ref{MPC1} LTV-MPC-LQR, \ref{LQR} LTI-MPC-LQR, all with horizon length ${N\!=\!1}$.
		\label{fig:impl_comparison}
	}
\end{figure*}

In Figure~\ref{fig:impl_comparison}, we show a comparison of the experimental results when using LTV-MPC-ADP, LTV-MPC-LQR and LTI-MPC-LQR, all with a horizon length of $N=1$. 
As can be seen in Figure~\ref{fig:impl_comparison}, the initial thrust is higher for LTV-MPC-ADP, which suggests that it is able to predict the behavior of the system better than LTV-MPC-LQR. It shows that for the same horizon length, LTV-MPC-ADP reacts faster in the correction of the drop in $p_z$, since the initial thrust input is larger. This suggests that using the ADP value function approximation for the final cost yields better predictions and therefore, performance.
The experimental comparison of LTV-MPC-ADP and LTI-MPC-LQR can be seen in the video at \cite{video}. For details on the implementation, we refer to \cite{dfallgit}.

\begin{figure}
	\centering
%
%
%
\definecolor{mycolor1}{rgb}{0.00000,0.44700,0.74100}%
\definecolor{mycolor2}{rgb}{0.85000,0.32500,0.09800}%

\begin{tikzpicture}

\begin{axis}[%
width=1.3in,
height=1.3in,
at={(0.884in,0.699in)},
scale only axis,
xmin=1,
xmax=11,
xlabel style={font=\color{white!15!black}, font=\footnotesize},
xlabel={Horizon length},
ymin=4000,
ymax=4500,
ylabel style={font=\color{white!15!black}, font=\footnotesize},
ylabel={$\sum{l(x_k,u_k)}$},
axis background/.style={fill=white},
xmajorgrids,
ymajorgrids,
scaled y ticks = true,
font=\footnotesize, 
ytick={4000, 4200, 4400},
legend style={at={(0.5,0.97)}, anchor=north, legend cell align=left, align=left, draw=white!15!black,, font=\footnotesize}
]
\addplot [color=mycolor1, line width=\lineWidth, mark=o, mark options={solid, mycolor1}]
  table[row sep=crcr]{%
1	4090.30380367308\\
2	4094.13726268001\\
3	4081.32645375361\\
4	4083.33828271879\\
5	4076.40896119517\\
6	4078.0484383517\\
7	4073.30454453154\\
8	4075.03309535723\\
9	4071.90446551784\\
10	4071.36755041019\\
11	4071.22557710747\\
12	4070.53646113455\\
};
\label{MPCADP_cost}

\addplot [color=black, dotted, line width=\lineWidth, mark=o, mark options={solid, black}]
  table[row sep=crcr]{%
1	4119.3854805619\\
2	4115.25120276441\\
3	4092.19983117712\\
4	4093.3912735374\\
5	4079.23868153477\\
6	4080.98783026224\\
7	4074.05115763685\\
8	4074.79805965544\\
9	4071.87223493769\\
10	4071.62309144344\\
11	4071.34365477832\\
12	4071.55055446306\\
};
\label{MPC_cost}

\addplot [color=mycolor2, dashed, line width=\lineWidth, mark=o, mark options={solid, mycolor2}]
table[row sep=crcr]{%
1 4449.965101293009\\
2 4448.564178972766\\
3 4447.972790497783\\
4 4448.822659128902\\
5 4451.168772241443\\
6 4452.826522368147\\
7 4454.925611510351\\
8 4456.414893208696\\
9 4458.352888391429\\
10 4462.135149148815\\
11 4466.249407503617\\
12 4467.998451360478\\
};\label{linear_MPC_PLQR_cost}

\end{axis}
\end{tikzpicture}%
	\caption{
		Costs for different horizon lengths, for policies: \ref{MPCADP_cost} LTV-MPC-ADP, \ref{MPC_cost}LTV-MPC-LQR and \ref{linear_MPC_PLQR_cost}LTI-MPC-LQR.
		\label{fig:costs_horizon}
	}
\end{figure}
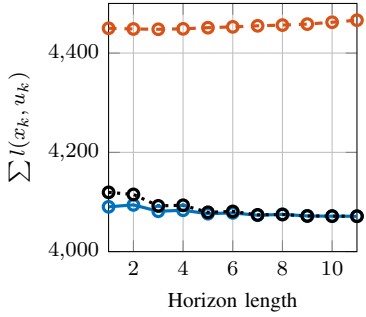

Figure~\ref{fig:costs_horizon} shows a series of simulations using the LTV-MPC-ADP, the LTV-MPC-LQR, and the LTI-MPC-LQR policies for horizon lengths from 1 up to 11.
For every horizon length, the sum of stage costs, $\sum_{k=0}^N l(x_k,u_k)$ is computed, until all states and inputs have converged, and then plotted as a measurement of the performance of the policy. As the aim is to minimize \eqref{eq:sum_stage_costs}, the smaller this cost is, the better the controller is performing.
Figure~\ref{fig:costs_horizon} supports the results of the performance comparison given in  Figure~\ref{fig:comparison_simulation}, as for short horizon lengths the cost  of LTV-MPC-ADP is lower than the one of LTV-MPC-LQR. It also shows that the quality of the approximation of the final cost loses importance when the horizon length increases.

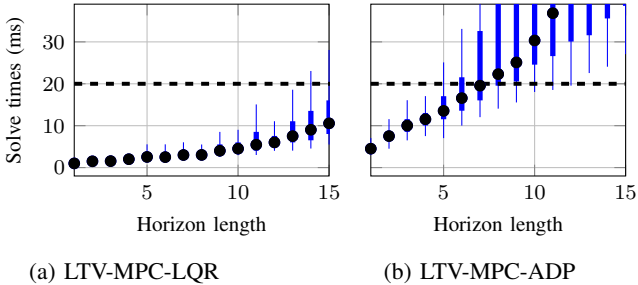
\begin{figure}
	%
	%
	\begin{subfigure}{0.19\textwidth}%
%
%

\begin{tikzpicture}
\begin{axis}[%
width=1.3337in,
height=0.90in,
at={(0.802in,0.79in)},
scale only axis,
clip=true,
xmin=1,
xmax=15,
xlabel style={font=\color{white!15!black}, font=\footnotesize},
xlabel={Horizon length},
ymin=-2,
ymax=39,
ylabel style={font=\color{white!15!black}, font=\footnotesize},
ylabel={Solve times (ms)},
axis background/.style={fill=white},
tick label style={/pgf/number format/fixed, font=\footnotesize},
xmajorgrids,
ymajorgrids,
legend style={legend cell align=left, align=left, draw=white!15!black, font=\footnotesize}
]
\addplot [color=blue, forget plot]
  table[row sep=crcr]{%
1	0.949859619140625\\
1	1.05094909667969\\
};
\addplot [color=blue, forget plot]
  table[row sep=crcr]{%
2	0.501632690429688\\
2	2.04658508300781\\
};
\addplot [color=blue, forget plot]
  table[row sep=crcr]{%
3	0.9918212890625\\
3	2.54440307617188\\
};
\addplot [color=blue, forget plot]
  table[row sep=crcr]{%
4	1.48582458496094\\
4	3.0517578125\\
};
\addplot [color=blue, forget plot]
  table[row sep=crcr]{%
5	1.50299072265625\\
5	5.55229187011719\\
};
\addplot [color=blue, forget plot]
  table[row sep=crcr]{%
6	1.50108337402344\\
6	5.55229187011719\\
};
\addplot [color=blue, forget plot]
  table[row sep=crcr]{%
7	1.98936462402344\\
7	6.04057312011719\\
};
\addplot [color=blue, forget plot]
  table[row sep=crcr]{%
8	2.00271606445313\\
8	5.51795959472656\\
};
\addplot [color=blue, forget plot]
  table[row sep=crcr]{%
9	2.54440307617188\\
9	8.52203369140625\\
};
\addplot [color=blue, forget plot]
  table[row sep=crcr]{%
10	2.96974182128906\\
10	9.02366638183594\\
};
\addplot [color=blue, forget plot]
  table[row sep=crcr]{%
11	3.00788879394531\\
11	15.07568359375\\
};
\addplot [color=blue, forget plot]
  table[row sep=crcr]{%
12	4.00733947753906\\
12	11.0664367675781\\
};
\addplot [color=blue, forget plot]
  table[row sep=crcr]{%
13	4.04739379882813\\
13	18.5832977294922\\
};
\addplot [color=blue, forget plot]
  table[row sep=crcr]{%
14	4.55093383789063\\
14	23.0579376220703\\
};
\addplot [color=blue, forget plot]
  table[row sep=crcr]{%
15	5.51414489746094\\
15	28.0723571777344\\
};
\addplot [color=blue, line width=2pt, forget plot]
  table[row sep=crcr]{%
1	0.997543334960938\\
1	1.02996826171875\\
};
\addplot [color=blue, line width=2pt, forget plot]
  table[row sep=crcr]{%
2	1.00326538085938\\
2	1.50680541992188\\
};
\addplot [color=blue, line width=2pt, forget plot]
  table[row sep=crcr]{%
3	1.50299072265625\\
3	2.01225280761719\\
};
\addplot [color=blue, line width=2pt, forget plot]
  table[row sep=crcr]{%
4	2.00271606445313\\
4	2.54440307617188\\
};
\addplot [color=blue, line width=2pt, forget plot]
  table[row sep=crcr]{%
5	2.05039978027344\\
5	3.50761413574219\\
};
\addplot [color=blue, line width=2pt, forget plot]
  table[row sep=crcr]{%
6	2.04277038574219\\
6	3.50761413574219\\
};
\addplot [color=blue, line width=2pt, forget plot]
  table[row sep=crcr]{%
7	2.50625610351563\\
7	4.00352478027344\\
};
\addplot [color=blue, line width=2pt, forget plot]
  table[row sep=crcr]{%
8	2.99072265625\\
8	4.01115417480469\\
};
\addplot [color=blue, line width=2pt, forget plot]
  table[row sep=crcr]{%
9	3.509521484375\\
9	5.54656982421875\\
};
\addplot [color=blue, line width=2pt, forget plot]
  table[row sep=crcr]{%
10	4.00924682617188\\
10	6.01959228515625\\
};
\addplot [color=blue, line width=2pt, forget plot]
  table[row sep=crcr]{%
11	4.04548645019531\\
11	8.52394104003906\\
};
\addplot [color=blue, line width=2pt, forget plot]
  table[row sep=crcr]{%
12	5.01251220703125\\
12	7.5531005859375\\
};
\addplot [color=blue, line width=2pt, forget plot]
  table[row sep=crcr]{%
13	6.01577758789063\\
13	11.0626220703125\\
};
\addplot [color=blue, line width=2pt, forget plot]
  table[row sep=crcr]{%
14	6.51741027832031\\
14	13.5345458984375\\
};
\addplot [color=blue, line width=2pt, forget plot]
  table[row sep=crcr]{%
15	8.02040100097656\\
15	16.0427093505859\\
};
\addplot [color=black, draw=none, mark=*, mark options={solid, fill=black, blue}, forget plot]
  table[row sep=crcr]{%
1	1.00326538085938\\
};
\addplot [color=black, draw=none, mark=*, mark options={solid, fill=black, blue}, forget plot]
  table[row sep=crcr]{%
2	1.50299072265625\\
};
\addplot [color=black, draw=none, mark=*, mark options={solid, fill=black, blue}, forget plot]
  table[row sep=crcr]{%
3	1.54781341552734\\
};
\addplot [color=black, draw=none, mark=*, mark options={solid, fill=black, blue}, forget plot]
  table[row sep=crcr]{%
4	2.01034545898438\\
};
\addplot [color=black, draw=none, mark=*, mark options={solid, fill=black, blue}, forget plot]
  table[row sep=crcr]{%
5	2.54631042480469\\
};
\addplot [color=black, draw=none, mark=*, mark options={solid, fill=black, blue}, forget plot]
  table[row sep=crcr]{%
6	2.50625610351563\\
};
\addplot [color=black, draw=none, mark=*, mark options={solid, fill=black, blue}, forget plot]
  table[row sep=crcr]{%
7	3.0059814453125\\
};
\addplot [color=black, draw=none, mark=*, mark options={solid, fill=black, blue}, forget plot]
  table[row sep=crcr]{%
8	3.02600860595703\\
};
\addplot [color=black, draw=none, mark=*, mark options={solid, fill=black, blue}, forget plot]
  table[row sep=crcr]{%
9	4.01687622070313\\
};
\addplot [color=black, draw=none, mark=*, mark options={solid, fill=black, blue}, forget plot]
  table[row sep=crcr]{%
10	4.51183319091797\\
};
\addplot [color=black, draw=none, mark=*, mark options={solid, fill=black, blue}, forget plot]
  table[row sep=crcr]{%
11	5.46455383300781\\
};
\addplot [color=black, draw=none, mark=*, mark options={solid, fill=black, blue}, forget plot]
  table[row sep=crcr]{%
12	6.01577758789063\\
};
\addplot [color=black, draw=none, mark=*, mark options={solid, fill=black, blue}, forget plot]
  table[row sep=crcr]{%
13	7.51304626464844\\
};
\addplot [color=black, draw=none, mark=*, mark options={solid, fill=black, blue}, forget plot]
  table[row sep=crcr]{%
14	9.02366638183594\\
};
\addplot [color=black, draw=none, mark=*, mark options={solid, fill=black, blue}, forget plot]
  table[row sep=crcr]{%
15	10.5628967285156\\
};
\addplot [color=black, draw=none, mark=*, mark options={solid, black}, forget plot]
  table[row sep=crcr]{%
1	1.00326538085938\\
};
\addplot [color=black, draw=none, mark=*, mark options={solid, black}, forget plot]
  table[row sep=crcr]{%
2	1.50299072265625\\
};
\addplot [color=black, draw=none, mark=*, mark options={solid, black}, forget plot]
  table[row sep=crcr]{%
3	1.54781341552734\\
};
\addplot [color=black, draw=none, mark=*, mark options={solid, black}, forget plot]
  table[row sep=crcr]{%
4	2.01034545898438\\
};
\addplot [color=black, draw=none, mark=*, mark options={solid, black}, forget plot]
  table[row sep=crcr]{%
5	2.54631042480469\\
};
\addplot [color=black, draw=none, mark=*, mark options={solid, black}, forget plot]
  table[row sep=crcr]{%
6	2.50625610351563\\
};
\addplot [color=black, draw=none, mark=*, mark options={solid, black}, forget plot]
  table[row sep=crcr]{%
7	3.0059814453125\\
};
\addplot [color=black, draw=none, mark=*, mark options={solid, black}, forget plot]
  table[row sep=crcr]{%
8	3.02600860595703\\
};
\addplot [color=black, draw=none, mark=*, mark options={solid, black}, forget plot]
  table[row sep=crcr]{%
9	4.01687622070313\\
};
\addplot [color=black, draw=none, mark=*, mark options={solid, black}, forget plot]
  table[row sep=crcr]{%
10	4.51183319091797\\
};
\addplot [color=black, draw=none, mark=*, mark options={solid, black}, forget plot]
  table[row sep=crcr]{%
11	5.46455383300781\\
};
\addplot [color=black, draw=none, mark=*, mark options={solid, black}, forget plot]
  table[row sep=crcr]{%
12	6.01577758789063\\
};
\addplot [color=black, draw=none, mark=*, mark options={solid, black}, forget plot]
  table[row sep=crcr]{%
13	7.51304626464844\\
};
\addplot [color=black, draw=none, mark=*, mark options={solid, black}, forget plot]
  table[row sep=crcr]{%
14	9.02366638183594\\
};
\addplot [color=black, draw=none, mark=*, mark options={solid, black}, forget plot]
  table[row sep=crcr]{%
15	10.5628967285156\\
};
\node[left, align=right, rotate=90]
at (0cm,-0.503cm) {1};
\node[left, align=right, rotate=90]
at (0.867cm,-0.503cm) {2};
\node[left, align=right, rotate=90]
at (1.734cm,-0.503cm) {3};
\node[left, align=right, rotate=90]
at (2.601cm,-0.503cm) {4};
\node[left, align=right, rotate=90]
at (3.468cm,-0.503cm) {5};
\node[left, align=right, rotate=90]
at (4.335cm,-0.503cm) {6};
\node[left, align=right, rotate=90]
at (5.202cm,-0.503cm) {7};
\node[left, align=right, rotate=90]
at (6.07cm,-0.503cm) {8};
\node[left, align=right, rotate=90]
at (6.937cm,-0.503cm) {9};
\node[left, align=right, rotate=90]
at (7.804cm,-0.503cm) {10};
\node[left, align=right, rotate=90]
at (8.671cm,-0.503cm) {11};
\node[left, align=right, rotate=90]
at (9.538cm,-0.503cm) {12};
\node[left, align=right, rotate=90]
at (10.405cm,-0.503cm) {13};
\node[left, align=right, rotate=90]
at (11.272cm,-0.503cm) {14};
\node[left, align=right, rotate=90]
at (12.139cm,-0.503cm) {15};
\addplot [color=black, dashed, line width=1.5pt]
  table[row sep=crcr]{%
1	20\\
20	20\\
};

\end{axis}
\end{tikzpicture}%
		\caption{LTV-MPC-LQR}
	\end{subfigure}
	\hspace{1.1cm}
	\begin{subfigure}{0.19\textwidth}%
%
%

\begin{tikzpicture}

\begin{axis}[%
width=1.3337in,
height=0.90in,
at={(0.802in,0.79in)},
scale only axis,
clip=true,
xmin=1,
xmax=15,
yticklabels={\empty},
xlabel style={font=\color{white!15!black}, font=\footnotesize},
xlabel={Horizon length},
ymin=-2,
ymax=39,
ylabel style={font=\color{white!15!black}, font=\footnotesize},
axis background/.style={fill=white},
tick label style={/pgf/number format/fixed, font=\footnotesize},
xmajorgrids,
ymajorgrids,
legend style={legend cell align=left, align=left, draw=white!15!black, font=\footnotesize}
]
\addplot [color=blue, forget plot]
  table[row sep=crcr]{%
1	2.98881530761719\\
1	7.05146789550781\\
};
\addplot [color=blue, forget plot]
  table[row sep=crcr]{%
2	4.50706481933594\\
2	11.5699768066406\\
};
\addplot [color=blue, forget plot]
  table[row sep=crcr]{%
3	6.500244140625\\
3	16.0808563232422\\
};
\addplot [color=blue, forget plot]
  table[row sep=crcr]{%
4	7.51876831054688\\
4	17.059326171875\\
};
\addplot [color=blue, forget plot]
  table[row sep=crcr]{%
5	7.03239440917969\\
5	25.1064300537109\\
};
\addplot [color=blue, forget plot]
  table[row sep=crcr]{%
6	10.0269317626953\\
6	33.0982208251953\\
};
\addplot [color=blue, forget plot]
  table[row sep=crcr]{%
7	12.0391845703125\\
7	56.6883087158203\\
};
\addplot [color=blue, forget plot]
  table[row sep=crcr]{%
8	14.0666961669922\\
8	84.2247009277344\\
};
\addplot [color=blue, forget plot]
  table[row sep=crcr]{%
9	15.5487060546875\\
9	86.273193359375\\
};
\addplot [color=blue, forget plot]
  table[row sep=crcr]{%
10	18.0473327636719\\
10	140.966415405273\\
};
\addplot [color=blue, forget plot]
  table[row sep=crcr]{%
11	18.5298919677734\\
11	162.973403930664\\
};
\addplot [color=blue, forget plot]
  table[row sep=crcr]{%
12	19.5560455322266\\
12	158.962249755859\\
};
\addplot [color=blue, forget plot]
  table[row sep=crcr]{%
13	22.5601196289063\\
13	181.018829345703\\
};
\addplot [color=blue, forget plot]
  table[row sep=crcr]{%
14	24.0764617919922\\
14	196.022033691406\\
};
\addplot [color=blue, forget plot]
  table[row sep=crcr]{%
15	27.069091796875\\
15	230.693817138672\\
};
\addplot [color=blue, line width=2pt, forget plot]
  table[row sep=crcr]{%
1	4.00924682617188\\
1	5.50079345703125\\
};
\addplot [color=blue, line width=2pt, forget plot]
  table[row sep=crcr]{%
2	6.5155029296875\\
2	8.5601806640625\\
};
\addplot [color=blue, line width=2pt, forget plot]
  table[row sep=crcr]{%
3	8.52394104003906\\
3	11.5699768066406\\
};
\addplot [color=blue, line width=2pt, forget plot]
  table[row sep=crcr]{%
4	10.040283203125\\
4	13.0348205566406\\
};
\addplot [color=blue, line width=2pt, forget plot]
  table[row sep=crcr]{%
5	11.5394592285156\\
5	17.0421600341797\\
};
\addplot [color=blue, line width=2pt, forget plot]
  table[row sep=crcr]{%
6	13.5746002197266\\
6	21.5568542480469\\
};
\addplot [color=blue, line width=2pt, forget plot]
  table[row sep=crcr]{%
7	16.0427093505859\\
7	32.5927734375\\
};
\addplot [color=blue, line width=2pt, forget plot]
  table[row sep=crcr]{%
8	19.5503234863281\\
8	45.6218719482422\\
};
\addplot [color=blue, line width=2pt, forget plot]
  table[row sep=crcr]{%
9	20.5497741699219\\
9	47.6360321044922\\
};
\addplot [color=blue, line width=2pt, forget plot]
  table[row sep=crcr]{%
10	24.5685577392578\\
10	72.1931457519531\\
};
\addplot [color=blue, line width=2pt, forget plot]
  table[row sep=crcr]{%
11	26.6056060791016\\
11	82.2181701660156\\
};
\addplot [color=blue, line width=2pt, forget plot]
  table[row sep=crcr]{%
12	30.0807952880859\\
12	91.2818908691406\\
};
\addplot [color=blue, line width=2pt, forget plot]
  table[row sep=crcr]{%
13	31.5837860107422\\
13	98.3028411865234\\
};
\addplot [color=blue, line width=2pt, forget plot]
  table[row sep=crcr]{%
14	35.5968475341797\\
14	107.311248779297\\
};
\addplot [color=blue, line width=2pt, forget plot]
  table[row sep=crcr]{%
15	38.6009216308594\\
15	124.832153320313\\
};
\addplot [color=black, draw=none, mark=*, mark options={solid, fill=black, blue}, forget plot]
  table[row sep=crcr]{%
1	4.51087951660156\\
};
\addplot [color=black, draw=none, mark=*, mark options={solid, fill=black, blue}, forget plot]
  table[row sep=crcr]{%
2	7.52067565917969\\
};
\addplot [color=black, draw=none, mark=*, mark options={solid, fill=black, blue}, forget plot]
  table[row sep=crcr]{%
3	10.0259780883789\\
};
\addplot [color=black, draw=none, mark=*, mark options={solid, fill=black, blue}, forget plot]
  table[row sep=crcr]{%
4	11.5394592285156\\
};
\addplot [color=black, draw=none, mark=*, mark options={solid, fill=black, blue}, forget plot]
  table[row sep=crcr]{%
5	13.5354995727539\\
};
\addplot [color=black, draw=none, mark=*, mark options={solid, fill=black, blue}, forget plot]
  table[row sep=crcr]{%
6	16.5824890136719\\
};
\addplot [color=black, draw=none, mark=*, mark options={solid, fill=black, blue}, forget plot]
  table[row sep=crcr]{%
7	19.5512771606445\\
};
\addplot [color=black, draw=none, mark=*, mark options={solid, fill=black, blue}, forget plot]
  table[row sep=crcr]{%
8	22.3064422607422\\
};
\addplot [color=black, draw=none, mark=*, mark options={solid, fill=black, blue}, forget plot]
  table[row sep=crcr]{%
9	25.1026153564453\\
};
\addplot [color=black, draw=none, mark=*, mark options={solid, fill=black, blue}, forget plot]
  table[row sep=crcr]{%
10	30.3354263305664\\
};
\addplot [color=black, draw=none, mark=*, mark options={solid, fill=black, blue}, forget plot]
  table[row sep=crcr]{%
11	36.8461608886719\\
};
\addplot [color=black, draw=none, mark=*, mark options={solid, fill=black, blue}, forget plot]
  table[row sep=crcr]{%
12	41.3703918457031\\
};
\addplot [color=black, draw=none, mark=*, mark options={solid, fill=black, blue}, forget plot]
  table[row sep=crcr]{%
13	45.6027984619141\\
};
\addplot [color=black, draw=none, mark=*, mark options={solid, fill=black, blue}, forget plot]
  table[row sep=crcr]{%
14	55.6478500366211\\
};
\addplot [color=black, draw=none, mark=*, mark options={solid, fill=black, blue}, forget plot]
  table[row sep=crcr]{%
15	65.6766891479492\\
};
\addplot [color=black, draw=none, mark=*, mark options={solid, black}, forget plot]
  table[row sep=crcr]{%
1	4.51087951660156\\
};
\addplot [color=black, draw=none, mark=*, mark options={solid, black}, forget plot]
  table[row sep=crcr]{%
2	7.52067565917969\\
};
\addplot [color=black, draw=none, mark=*, mark options={solid, black}, forget plot]
  table[row sep=crcr]{%
3	10.0259780883789\\
};
\addplot [color=black, draw=none, mark=*, mark options={solid, black}, forget plot]
  table[row sep=crcr]{%
4	11.5394592285156\\
};
\addplot [color=black, draw=none, mark=*, mark options={solid, black}, forget plot]
  table[row sep=crcr]{%
5	13.5354995727539\\
};
\addplot [color=black, draw=none, mark=*, mark options={solid, black}, forget plot]
  table[row sep=crcr]{%
6	16.5824890136719\\
};
\addplot [color=black, draw=none, mark=*, mark options={solid, black}, forget plot]
  table[row sep=crcr]{%
7	19.5512771606445\\
};
\addplot [color=black, draw=none, mark=*, mark options={solid, black}, forget plot]
  table[row sep=crcr]{%
8	22.3064422607422\\
};
\addplot [color=black, draw=none, mark=*, mark options={solid, black}, forget plot]
  table[row sep=crcr]{%
9	25.1026153564453\\
};
\addplot [color=black, draw=none, mark=*, mark options={solid, black}, forget plot]
  table[row sep=crcr]{%
10	30.3354263305664\\
};
\addplot [color=black, draw=none, mark=*, mark options={solid, black}, forget plot]
  table[row sep=crcr]{%
11	36.8461608886719\\
};
\addplot [color=black, draw=none, mark=*, mark options={solid, black}, forget plot]
  table[row sep=crcr]{%
12	41.3703918457031\\
};
\addplot [color=black, draw=none, mark=*, mark options={solid, black}, forget plot]
  table[row sep=crcr]{%
13	45.6027984619141\\
};
\addplot [color=black, draw=none, mark=*, mark options={solid, black}, forget plot]
  table[row sep=crcr]{%
14	55.6478500366211\\
};
\addplot [color=black, draw=none, mark=*, mark options={solid, black}, forget plot]
  table[row sep=crcr]{%
15	65.6766891479492\\
};
\node[left, align=right, rotate=90]
at (0cm,-0.503cm) {1};
\node[left, align=right, rotate=90]
at (0.867cm,-0.503cm) {2};
\node[left, align=right, rotate=90]
at (1.734cm,-0.503cm) {3};
\node[left, align=right, rotate=90]
at (2.601cm,-0.503cm) {4};
\node[left, align=right, rotate=90]
at (3.468cm,-0.503cm) {5};
\node[left, align=right, rotate=90]
at (4.335cm,-0.503cm) {6};
\node[left, align=right, rotate=90]
at (5.202cm,-0.503cm) {7};
\node[left, align=right, rotate=90]
at (6.07cm,-0.503cm) {8};
\node[left, align=right, rotate=90]
at (6.937cm,-0.503cm) {9};
\node[left, align=right, rotate=90]
at (7.804cm,-0.503cm) {10};
\node[left, align=right, rotate=90]
at (8.671cm,-0.503cm) {11};
\node[left, align=right, rotate=90]
at (9.538cm,-0.503cm) {12};
\node[left, align=right, rotate=90]
at (10.405cm,-0.503cm) {13};
\node[left, align=right, rotate=90]
at (11.272cm,-0.503cm) {14};
\node[left, align=right, rotate=90]
at (12.139cm,-0.503cm) {15};
\addplot [color=black, dashed, line width=1.5pt]
  table[row sep=crcr]{%
1	20\\
20	20\\
};

\end{axis}
\end{tikzpicture}%
		\caption{LTV-MPC-ADP}
	\end{subfigure}
	\caption{
		Comparison of the solve times reported by the Gurobi solver \cite{gurobi} when playing the respective policy online. The black dotted line show the 50\,Hz real-time limit.
		\label{fig:comparison_solve_times}
	}
\end{figure}


Figure~\ref{fig:comparison_solve_times} shows that the maximum implementable horizon length for the LTV-MPC-ADP policy at 50\,Hz is of 4-5 steps, while the LTV-MPC-LQR policy is tractable for up to 15 steps. 
The online computation time of the LTV-MPC-ADP policy can be reduced by using fewer approximate value functions in $V_{PWM}^*$, which introduces fewer quadratic constraints in \eqref{eq:optimization_problem_general_ADP}. This generally introduces a trade-off with the online performance because the approximation quality of the cost-to-go may get reduced. Moreover, reducing the computation time allows for a higher frequency which might improve the forward Euler approximation used for the discrete time dynamics. 

%
The offline computation time required to compute the policies LTV-MPC-LQR and LTI-MPC-LQR is solving the Riccati equation, which takes less than one second for this model. The offline computation required to prepare $\hat{V}^*_{PWM}$ for the LTV-MPC-ADP policy is less than 10 minutes on a standard desktop computer, using Yalmip \cite{lofberg_2004_yalmip} and Mosek \cite{mosek} for building and solving the necessary SDPs.


\section{Conclusion and Future Work}

We applied Approximate Dynamic Programming methods to a nonlinear, high dimensional quadcopter system with continuous state and input spaces. 
Using polynomial approximations of the dynamics, the constraints and the value function, leads to an online policy capturing the relevant nonlinearities of the system. 
For computational tractability, the ADP method was combined with an MPC scheme, which leverages the computational benefits of a short-term MPC scheme and the long term advantages of capturing the effects of nonlinearities by the ADP value function.  
The latter was used as the terminal cost in a linear time-varying MPC. For short time horizons this control method outperformed the same MPC with the terminal cost being the solution to a Riccati equation.  
The performance of the resulting controller was successfully demonstrated both in simulations and in experiments on a nano-quadcopter.

In future work, the control performance could further be improved by allowing for higher order polynomial approximations of the dynamics, or of  relevant parts of the dynamics while keeping the complexity of other parts low, in order to ensure computational tractability.   
Furthermore, formulating the dynamics in terms of quaternions instead of Euler angles is promising as quaternion dynamics are naturally polynomial. However, more investigation is needed on how to effectively approximate the value function that only takes values on the quaternion manifold.

\addtolength{\textheight}{-0cm}   




\bibliographystyle{ieeetr}
\bibliography{07_bib}


%
\end{document}